\documentclass[preprint,12pt]{elsarticle}

\usepackage{amssymb}

\usepackage{amsthm}
\newtheorem{Def}{Definition}

\usepackage{amsmath}
\usepackage{amssymb}
\usepackage{multirow}
\usepackage{booktabs}
\usepackage{graphicx}
\usepackage{float}
\usepackage{subfig}
\usepackage{longtable}

\usepackage{tabu}                     % 表格插入
\usepackage{multirow}                 % 一般用以设计表格，将所行合并
\usepackage{multicol}                 % 合并多列
\usepackage{multirow}                % 合并多行
\usepackage{float}                    % 图片浮动
\usepackage{makecell}                 % 三线表-竖线
\usepackage{booktabs}                 % 三线表-短细横线

\makeatletter
\def\@author#1{\g@addto@macro\elsauthors{\normalsize%
    \def\baselinestretch{1}%
    \upshape\authorsep#1\unskip\textsuperscript{%
      \ifx\@fnmark\@empty\else\unskip\sep\@fnmark\let\sep=,\fi
      \ifx\@corref\@empty\else\unskip\sep\@corref\let\sep=,\fi
      }%
    \def\authorsep{\unskip,\space}%
    \global\let\@fnmark\@empty
    \global\let\@corref\@empty  %% Added
    \global\let\sep\@empty}%
    \@eadauthor={#1}
}
\makeatother

\usepackage{soul}
\usepackage{color, xcolor}


\usepackage{algorithm}
\usepackage{algorithmicx}
\usepackage{algpseudocode}

\usepackage{bbm}
\usepackage{mathrsfs}

\usepackage{setspace}

\journal{IEEE Transactions on Cybernetics}

\begin{document}

\begin{frontmatter}
%% Title, authors and addresses

%% use the tnoteref command within \title for footnotes;
%% use the tnotetext command for theassociated footnote;
%% use the fnref command within \author or \address for footnotes;
%% use the fntext command for theassociated footnote;
%% use the corref command within \author for corresponding author footnotes;
%% use the cortext command for theassociated footnote;
%% use the ead command for the email address,
%% and the form \ead[url] for the home page:
%% \title{Title\tnoteref{label1}}
%% \tnotetext[label1]{}
%% \author{Name\corref{cor1}\fnref{label2}}
%% \ead{email address}
%% \ead[url]{home page}
%% \fntext[label2]{}
%% \cortext[cor1]{}
%% \affiliation{organization={},
%%             addressline={},
%%             city={},
%%             postcode={},
%%             state={},
%%             country={}}
%% \fntext[label3]{}

\title{A New Quantum Dempster Rule of Combination}

%% use optional labels to link authors explicitly to addresses:
%% \author[label1,label2]{}
%% \affiliation[label1]{organization={},
%%             addressline={},
%%             city={},
%%             postcode={},
%%             state={},
%%             country={}}
%%
%% \affiliation[label2]{organization={},
%%             addressline={},
%%             city={},
%%             postcode={},
%%             state={},
%%             country={}}

%\author{Huaping He}
%\author{Fuyuan Xiao}

%\address{organization={},%Department and Organization
%            addressline={}, 
%            city={},
%            postcode={},
%            state={},
%            country={}}
\author[label2]{Huaping He}
%\ead{hehuapingswu@163.com}

\author[label2]{Fuyuan Xiao\corref{cor1}}
\ead{doctorxiaofy@hotmail.com; xiaofuyuan@cqu.edu.cn}

\cortext[cor1]{Corresponding author: Fuyuan Xiao, School of Big Data and Software Engineering, Chongqing University, Chongqing 401331, China.}

\address[label2]{School of Big Data and Software Engineering, Chongqing University, Chongqing, China, 401331.}
%\address[label2]{School of Computer and information science, Southwest University, Chongqing, China, 400715.}

\begin{abstract}
%% Text of abstract
Dempster rule of combination (DRC) is widely used for uncertainty reasoning in intelligent information system, which is generalized to complex domain recently. However, as the increase of identification framework elements, the computational complexity of  Dempster Rule of Combination increases exponentially. To address this issue, we propose a  novel quantum Dempster rule of combination (QDRC) by means of Toffoli gate. The QDRC combination process is completely implemented using quantum circuits.

%We conducted simulation experiments on IBM and IonQ quantum cloud platforms to verify that our algorithm is reasonable. At the same time, we also uploaded the quantum circuit to the real quantum computer of IonQ Harmony, the experiment shows that QDRC decision results will be more accurate as the number of measurements increases. Finally, we apply the algorithm to several groups of real UCI data classification.
\end{abstract}

%%Graphical abstract
%\begin{graphicalabstract}
%%\includegraphics{grabs}
%\end{graphicalabstract}
%
%%%Research highlights
%\begin{highlights}
%\item Research highlight 1
%\item Research highlight 2
%\end{highlights}

\begin{keyword}
Dempster rule of combination\sep Complex  Dempster Rule of Combination, \sep Quantum algorithm\sep Toffoli gate Quantum multisource information fusion

%% keywords here, in the form: keyword \sep keyword

%% PACS codes here, in the form: \PACS code \sep code

%% MSC codes here, in the form: \MSC code \sep code
%% or \MSC[2008] code \sep code (2000 is the default)

\end{keyword}
\end{frontmatter}

%% \linenumbers

%% main text
\section{Introduction}

 Compared with the classical computer, the quantum computer has a faster running speed, for which the quantum evidence theory is proposed \cite{Xiao2022GQET}. As the increase of identification framework elements, the computational complexity of Dempster Rule of Combination increases exponentially. For this reason, the quantum algorithm of Dempster rule of combination (QADRC) was propose \cite{2020Quantum}. In this rule, the quantum computer performs the calculation combination process, and the fusion time of the identification frame with multiple elements is significantly reduced compared with the classical computer. However, QDRC only partially implements the DRC process in quantum computers. It does not implement the combination steps in the classic DRC and cannot calculate the final decision result. In this case, we propose a novel quantum Dempster rule of combination (QDRC) by means of Toffoli gate. QDRC can be completely reproduced DRC in the quantum algorithm.\par
The main contributions are listed as follows.\par
\begin{itemize}

\item{The proposed QDRC is proposed on the based of Toffoli gate to solve as the increase of identification framework elements, the computational complexity of  Dempster Rule of Combination increases exponentially problem.}
\item{The proposed QDRC completely runs DRC in Quantum computer.}
\item{The QDRC is applied for multisource information in the IonQ.}

%\item{The QDRC algorithm proposed in this work uses the controlled gate to skillfully solve the logic operation required by the quantum circuit when combining evidence.}
%\item{The QDRC algorithm proposed in this paper completely realizes the function of DRC. The quantum computer can directly make decisions after using the quantum circuit in this paper, reducing the steps of classical computer combination in QADRC.}
%\item{This work runs the information fusion algorithm in a real quantum computer IonQ for the first time, and verifies the rationality of the algorithm. This is helpful to the popularization of DSET in quantum computers and to improve the ability of quantum computers to process uncertain information.}
\end{itemize}
The rest of the paper is organized as follows. The Section 2 will introduce the relevant knowledge of evidence theory and QADRC. The QDRC algorithm proposed in this paper will be introduced in the Section 3. In the Section 4, we will analyze and compare the combination method.  At last, We summarized our work in Section 5.

\section{Preliminaries}
	In this section, we will introduce the complex evidence theory (CET) and quantum algorithm of Dempster rule of combination (QADRC).
	
\subsection{Complex evidence theory}
Classical evidence theory is an inexact reasoning theory with the ability to handle uncertain information. It was first applied in multisource information fusion \cite{wang2022modelling,li2021multisource,liu2022consistency},
 expert systems \cite{chang2021transparent}, confidence rule libraries \cite{yager2011fusion, yager2019uncertain} and pattern recognition \cite{Xiao2022NQMF, Xiao2022Acomplexweighted,Shang2021compound}. The complex evidence theory can be applied to the processing of complex fuzzy sets. In addition, when the phase or period of Uncertain data changes, the complex evidence theory can reduce the impact of abnormal evidence in the process of processing. When the imaginary positions of each element of the mass function are equal to 0, the complex evidence theory degenerates into classical evidence theory.

\subsubsection{Quantum gate}
In quantum computing, quantum gates are the basis of quantum circuits. It is like the relationship between traditional logic gates and general digital circuits. Quantum gates generally operate on one or two bits, and can be represented by unitary matrices. Common quantum gates are shown in Table \ref{tab:Comparison}.

\begin{table}[!ht]
\footnotesize
\centering
% table caption is above the table
\caption{Basic quantum gate}
\label{tab:Comparison}       % Give a unique label
% For LaTeX tables use
\begin{tabular}{ccc}
\toprule
Gates & Notation & Matrix\\
\hline
Pauli-X gate & \makecell[c]{ \ \\
\begin{minipage}[b]{0.17\columnwidth}
    \centering
    {\includegraphics[width=0.9\textwidth]{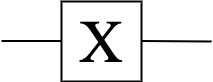}}
\end{minipage}\\ 
\
} & $\left[    
		\begin{array}{cc}
			0&1\\
			1&0
		\end{array} 		
		    \right] $   \\

Y-Rotation gate & \makecell[c]{\ \\
\begin{minipage}[b]{0.17\columnwidth}
    \centering
    {\includegraphics[width=0.9\textwidth]{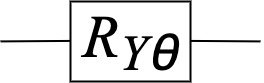}}
\end{minipage}\\
\
}&$\left[
		\begin{array}{cc}
			cos(\frac{\theta}{2})&$-$sin(\frac{\theta}{2})\\
			sin(\frac{\theta}{2})&cos(\frac{\theta}{2})
		\end{array}
			\right]	$ \\

CNOT gate & \makecell[c]{\ \\
\begin{minipage}[b]{0.3\columnwidth}
    \centering
    {\includegraphics[width=0.9\textwidth]{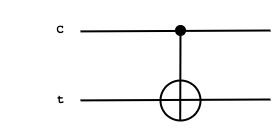}}
\end{minipage}\\
\
}& $	\left[
		\begin{array}{cccc}
			1 & 0 & 0 & 0\\
			0 & 1 & 0 & 0\\
			0 & 0 & 0 & 1\\
			0 & 0 & 1 & 0
		\end{array}
			\right]	$ \\

CNOT-Y-Rotation gate & \makecell[c]{\ \\
\begin{minipage}[b]{0.3\columnwidth}
    \centering
    {\includegraphics[width=0.9\textwidth]{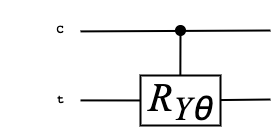}}
\end{minipage}\\
\
}& $	\left[
		\begin{array}{cccc}
			1 & 0 & 0 & 0\\
			0 & 1 & 0 & 0\\
			0 & 0 & cos(\frac{\theta}{2})&$-$sin(\frac{\theta}{2})\\
			0 & 0 & sin(\frac{\theta}{2})&cos(\frac{\theta}{2})
		\end{array}
			\right]	$\\

\hline

%Weight  & 0.8kg & 1kg & 1.3kg \\
%Vision  & Linear   & Linear  & Hybrid\\
%Function & \makecell[c]{Model\\Size}& \makecell[c]{Training\\Time}&\makecell[c]{Detection\\Time} \\
%Function  & 6   & 7  & 8\\
%Price & 2& 3 & 4\\
\hline
\end{tabular}
\end{table}

\subsubsection{Toffoli gate}
Quantum bits can be used to express information, and quantum circuits and quantum gates are another key point of quantum computing. A quantum computer is constructed by a quantum circuit containing a circuit and a basic quantum gate. Now we will focus on the Toffoli gate.\par

\begin{Def}(Toffoli gate)\par

	Toffoli gate is a three-qubit gate. Assume the bases state is
	$|000\rangle$, $|001\rangle$, $010\rangle$, $|011\rangle$, $|100\rangle$, $|101\rangle$, $|110\rangle$, $|111\rangle$. The unitary transformation performed by Toffoli gate can be defined as:
	
	\begin{equation}\label{Equation2.2_2}
		U_{CN}=\left[    
		\begin{array}{cccccccc}
			1 & 0 & 0 & 0 & 0 & 0 & 0 & 0\\
			0 & 1 & 0 & 0 & 0 & 0 & 0 & 0\\
			0 & 0 & 1 & 0 & 0 & 0 & 0 & 0\\
			0 & 0 & 0 & 1 & 0 & 0 & 0 & 0\\
			0 & 0 & 0 & 0 & 1 & 0 & 0 & 0\\
			0 & 1 & 0 & 0 & 0 & 1 & 0 & 0\\
			0 & 1 & 0 & 0 & 0 & 0 & 0 & 1\\
			0 & 1 & 0 & 0 & 0 & 0 & 1 & 0
		\end{array} 		
		    \right].
	\end{equation}	
	
	$U_{CN}$ is a unitary matrix, that is $U_{CN}^{\dagger}U_{CN}=I$. 
	
\end{Def}

\section{Analyzing of quantum algorithm of Dempster rule of combination}
\begin{figure}[h]\centering
		\includegraphics[width=0.95\textwidth]{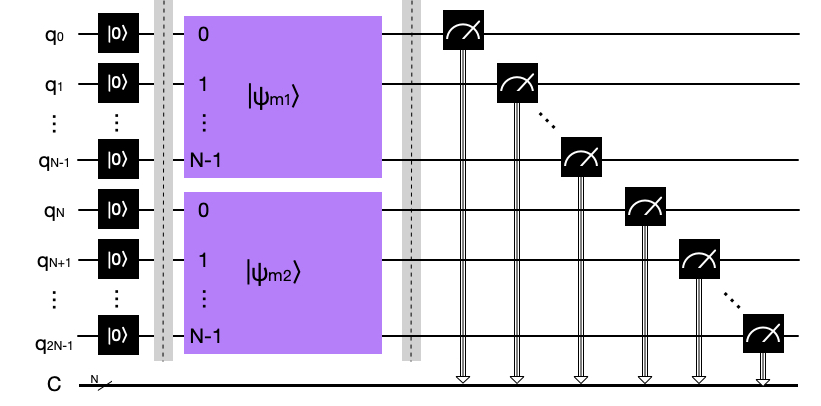}
		\caption{ The quantum circuit of QADRC. }
		\label{QADRC}
	\end{figure}

The quantum algorithm of Dempster rule of combination (QADRC) was proposed to solve the problem that the combination time of multi-element identification framework is slow \cite{2020Quantum}. It put the tensor calculation part of DRC into the quantum computer for the first time, and simulated the algorithm on the IBM quantum platform. From its experiment, we can see that the number of elements in the identification framework has little effect on the algorithm operation time. The quantum circuit diagram of QADRC is shown in Fig. \ref{QADRC}. $2N$ qubits are needed in the circuit, of which the qubits labeled $0$ to $N-1$ are used to represent $|\mathbbm{M}_1\rangle$, and the qubits labeled $N$ to $2N-1$ are used to represent $|\mathbbm{M}_2\rangle$. Finally, the probability of $|\mathbbm{M}_1\mathbbm{M}_2\rangle$ obtained by measuring these $2N$ qubits.

 However, QADRC only calculates two different BBA elements, and does not fully implement the DRC function in quantum computer environment. After measuring the $|\mathbbm{M}_1\mathbbm{M}_2\rangle$ result, it is still necessary to perform $A_i\bigcap A_j = A$ or $A_i\bigcap A_j = \varnothing$ in DRC, so as to eliminate the case that all quantum bits of $|\mathbbm{M}_1\rangle$ and $|\mathbbm{M}_2\rangle$ are mutually exclusive.

%第三节
\section{A new method for quantum  Dempster Rule of Combination  
Dempster rule of combination}

In order to solve the problem that QADRC needs the logic. In this section, we propose a new quantum Dempster rule of combination based on Toffoli gate (QDRC), which can be fully implemented in quantum computer.

%In this section, we will introduce the QDRC method. Fig. \ref{QDRC} shows the process of QDRC. After obtaining BBA or CBBA in a classical computer, the generation, combination and measurement of QBBA will be performed in a quantum computer. The steps of QDRC will be introduced in section 4.1, the quantum circuit of QDRC will be introduced in section 4.2, and the pseudocode of QDRC will be given in Section 4.3.
	
\subsection{QDRC algorithm}
In this section, we propose the technological process of QDRC. In the QDRC algorithm, the first step is data initialization. When the initial data is obtained, the BBA and CBBA are calculated by the classical computer. This is to obtain a mass function that is easy to express in qubits. The second step is  to initialize the quantum circuit and generate QBBA. The third step is to use Toffoli gate to carry out BBA/CBBA combination in quantum circuit. The fourth step is quantum measurement to get the combined results. Steps 2 to 4 completely run the whole DSET process in quantum computer.

\begin{itemize}
\item{Step 1: Data initialization}\par
	In quantum theory, the probability is expressed by the square of the module. In order to further expand the theory, the CBBA/BBA  is processed.
\begin{Def}($\mathcal{P}$ transform function)\par
	Assume there is a CBBA $\mathbbm{M}(A)$. The $\mathcal{P}$ transform function is defined as:
	\begin{equation}\label{eq4.1}
	 	\mathcal{P}_i = \sqrt{ \frac{ \mathcal{M}(A_i)}{\sum\limits_{i=1}^{2^N} \mathcal{M}(A_i)}},                 
	\end{equation}
	where\[ \mathcal{M}(A) =\left| \mathbbm{M} (A)\right| =\sqrt{(\mathcal{R}( \mathbbm{M} (A)))^2+(\mathcal{I}( \mathbbm{M} (A)))^2}, \]
	and $A_i$ is the $i$th element of the power set which based on the frame of discernment; $\mathcal{R}(\mathbbm{M}(\cdot))$ means to calculate the real part of $\mathbbm{M}(\cdot)$; $\mathcal{I}(\mathbbm{M}(\cdot))$ means to calculate the imaginary part of $\mathbbm{M}(\cdot)$; $\mathcal{P}_i$ satisfies $\sum \left|\mathcal{P}_i\right|^2=1$.

%	We use $real(\cdot)$ to represent the real part of the complex number and $imag(\cdot)$ to represent the imaginary part of the complex number. Then the GBBA after initialization is expressed as :
		
%	\begin{equation}\label{eq4.1}
%		\mathcal{M}(A)=\left| \mathbbm{M} (A)\right|=\sqrt{real( \mathbbm{M} (A))^2+imag( \mathbbm{M} (A))^2}.
%	\end{equation}
	
\end{Def}

	In the quantum circuit, the converted $\mathcal{P}$ represents the amplitude of the quantum bit, $\left | \mathcal{P}_i \right|^2$ represents the probability of the quantum bit. When all of $\mathcal{I}( \mathbbm{M}(\cdot))=0$, the CBBA $\mathbbm{M}$ degenerate to a BBA, BBA is expressed by $m$. In this case, the $\mathcal{P}$ conversion function can be expressed as:
	\begin{equation}\label{eq4.1}
		\mathcal{P}_i= m(A) .
	\end{equation}

\item{Step 2: Data encoding}\par
	Generally, in quantum machine learning, the methods of encoding data include basic encoding, amplitude encoding, angle encoding and arbitrary encoding. Basic encoding is very easy to understand, but the state vector of its encoding result may become very sparse. Angle encoding and arbitrary encoding only encodes one data point at a time, not the entire data set. The efficiency of these three encoding methods is not high. The amplitude coding to be used in this work requires only $\log_2{2^N}=N$ qubits for encoding each $\mathcal{M}$.
	
	\begin{Def}(QBBA generation function)\par
		Let $\Theta=\{\theta_1,\theta_2, \cdots, \theta_N \}$ be a frame of discernment. A QBBA is defined as:
		\begin{equation}\label{eq4.2}
			|\psi_\mathbbm{M}\rangle =\sum\limits_{i=1}^{2^N} \mathcal{P}_i | \mathcal{F}_i\rangle,   
		\end{equation}		
		where $\mathcal{F}_i$ is the qubits of $A_i$, they have a one-to-one correspondence.\par

	\end{Def}	\par

%	 In QDRC, we will use amplitude coding, because it can better express the probability of each proposition of the QBBA clock.
	
%	We use multiple qubits to express QBBA. A quantum bit represents a mutually exclusive event. For example, in Framework $\Theta=\{\theta_1,\theta_2, \cdots, \theta_N \}$, QBBA is expressed as:
%	\begin{equation}
		%|\psi\rangle=\alpha_{00}|00\rangle+\alpha_{01}|01\rangle+\alpha_{10}|10\rangle+\alpha_{11}|11\rangle,
%		|\psi_\mathbb{M}\rangle=\sum\limits_{i=1}^{2^N} p_i|F_i\rangle,
%	\end{equation}
%	where $p_i$ is the amplitude, it is a complex number. Using $\left| p_i \right|^2$ to express the support of proposition $F_i$. $F_i$ has a one-to-one correspondence with the elements in the power set $2^\Theta$, and meet $\sum \left|p_i\right|^2=1$.\par

	The amplitude encoding method embeds the classical data into the amplitude of the quantum state. In Qiskit language, we usually use $initialize(\cdot )$ function to generate QBBA.
	%In particular, if there are N element frames A, N qubits are required to initialize the QBBA circuit.\par

%	For example, under Framework $\Theta=\{\theta_1,\theta_2 \}$, there is BBA: $m=[\frac{1}{3}, \frac{1}{3}, \frac{1}{3}]$. Then QBBA is expressed as follows:
%	\begin{equation}
%		\psi_\mathbb{M}\rangle=0|00\rangle+\sqrt{\frac{1}{3}}|01\rangle+\sqrt{\frac{1}{3}}|10\rangle+\sqrt{\frac{1}{3}}|11\rangle.
%	\end{equation}

\item{Step 3: QBBA combination}\par
	Suppose there are two QBBA $|\psi_{\mathbbm{M}1}\rangle$ and $|\psi_{\mathbbm{M}2}\rangle$. The combined QBBA of $|\psi_{\mathbbm{M}1}\rangle$ and $|\psi_{\mathbbm{M}2}\rangle$ is denoted as $|\psi_{\mathbbm{M}}\rangle=|\psi_{\mathbbm{M}1}\rangle \otimes |\psi_{\mathbbm{M}2}\rangle$. The combination step consists of two parts: 1) Calculating the tensor product, and 2) Logical combination. In DSET, the tensor product calculation will complete the $\mathbbm{M}_1(A_i)\mathbbm{M}_2(A_i)$ and the logical combination is operated by $A_i \bigcap A_j = A $. In QDRC, by measuring two QBBAs, the combination result can be obtained and the tensor product can be obtained, while the logic judgment is completed through the Toffoli gate.
	
\item{Step 4: Combined result measurement}\par
	The system state can be obtained by measuring operator ${M_\mathcal{A}}$. The index $\mathcal{A}$ represents the results that may occur in the experiment. Measure the combined QBBA $|\psi_\mathbb{M}\rangle$, then the probability of $\mathbbm{M}$ is:
	\begin{equation}
		p(\mathcal{M} )=\langle \psi_\mathbb{M}| M_{\mathcal{A}}^{\dagger} M_\mathcal{A} |\psi_\mathbb{M} \rangle.
	\end{equation}
	In the third step of logical combination, we use a Toffoli gate controlled by 2-qubits, and will change the quantum bit $|0\rangle$ of the event to $|1\rangle$ when both $| \mathcal{F}_i \rangle$ and $|\mathcal{F}_j\rangle$ support the same event at the same time. However, when elements $|\mathcal{F}_i\rangle$ and $|\mathcal{F}_j\rangle$ are excluded, the QBBA of the combination result will not be affected, and the QBBA at this time will support empty sets ($|00\dots0\rangle$). This is because the input quantum bit defaults to $|00\dots0\rangle$, so we can ignore $|00\dots0\rangle$'s measurement results.
\end{itemize}

	%In a quantum circuit, the initialization of a quantum bit is always A, which cannot represent QBBA. We need to use a circuit to convert the initialized quantum into a superposed QBBA.
%	After quantum measurement, we can get the probability $x(=00,01,10,11)$, mark as $\left| \alpha_x \right|^2$. QBBA is generated by quantum circuit. Because of the initial state of the quantum circuit, the quantum bit is $|0\rangle$. 	This is not a superposition state. 	

\subsection{QDRC Quantum circuit}
\begin{figure}[htbt]\centering
		\includegraphics[width=1\textwidth]{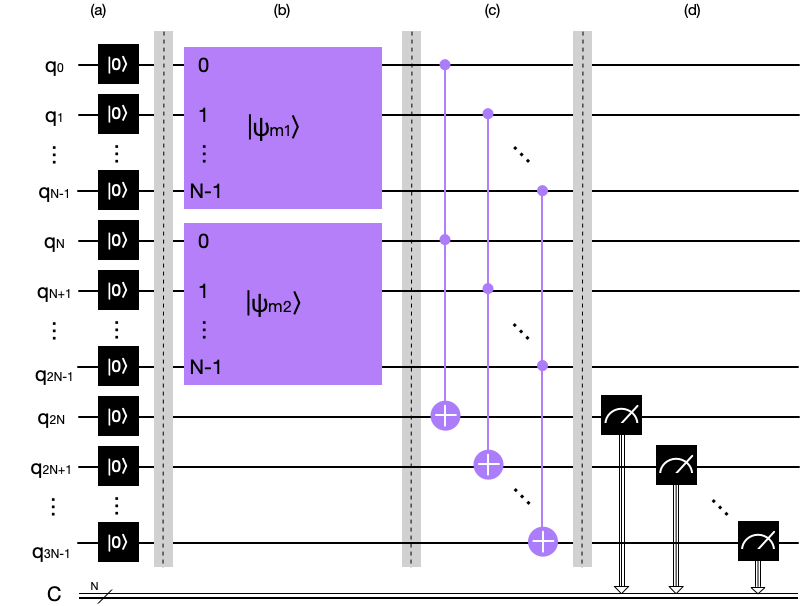}
		\caption{The Quantum circuit of QDRC. (a) Quantum circuit initialization. (b) Generation of QBBA in quantum circuits. (c) Quantum state combination (d) Result measurement}
		\label{dianlu}
	\end{figure}

	In QDRC, the above steps 2-4 are carried out in a quantum computer system. A quantum computer is a device that uses quantum logic for general computing. Usually we use quantum circuits to express quantum logic. In this subsection, we propose the framework of QDRC quantum circuit (Fig. \ref{dianlu}) and explain the role of each part of the circuit. Fig. \ref{dianlu}(a) is the initialization of quantum circuit operation. Fig. \ref{dianlu}(b) is the data encoding operation with quantum circuits, it can produce the QBBA. Fig. \ref{dianlu}(c) is the quantum state combination, it is the key step of QDRC, which combines two QBBAs in a quantum circuit. Fig. \ref{dianlu}(d) is the result measurement.

\section{Compare QDRC with other existing combination schemes}

The whole process of the special DRC and CDRC algorithms are calculated by the classical computer; QADRC uses quantum computers to calculate tensor products and classical computers to calculate logical combinations. Unlike the existing algorithms, the whole process of QDRC algorithm proposed in this paper is implemented in quantum computer. After measurement, the decision can be made directly. Because the Toffoli gate will not change the mutually exclusive qubit gate, and the initial value of qubit is 0. We can use $|00 \cdots 0 \rangle$ to $|11 \cdots 1 \rangle$ to calculate the numerical result.

%In the following part, we will run QDRC algorithm in IBM and IonQ quantum simulators and actual hardware, and discuss the combination results of several algorithms.

\section{Conclution}

With the intensification of the process of information diversification, the efficiency of information fusion has been put on the agenda. Quantum computers have powerful quantum information processing capabilities. It is an unsolved problem to rely on quantum computers to realize DRC. The QDRC algorithm proposed in this paper makes up for the deficiency that the classical computer still needs to carry out logical operation to obtain the results after the tensor is calculated by QDRC. For the first time, the whole process of DRC is deployed in a quantum computer.\par
QDRC will further strengthen the application of information fusion under the multi-element identification framework in the future work, develop a simpler learning mechanism, and expect to establish an appropriate quantum information fusion method to deal with various uncertain data.

%% The Appendices part is started with the command \appendix;
%% appendix sections are then done as normal sections
%% \appendix

%% \section{}
%% \label{}

%% For citations use: 
%%       \citet{<label>} ==> Jones et al. [21]
%%       \citep{<label>} ==> [21]
%%

%% If you have bibdatabase file and want bibtex to generate the
%% bibitems, please use
%%
%%  \bibliographystyle{elsarticle-num-names} 
%%  \bibliography{<your bibdatabase>}

\begin{thebibliography}{10}
\expandafter\ifx\csname url\endcsname\relax
  \def\url#1{\texttt{#1}}\fi
\expandafter\ifx\csname urlprefix\endcsname\relax\def\urlprefix{URL }\fi
\expandafter\ifx\csname href\endcsname\relax
  \def\href#1#2{#2} \def\path#1{#1}\fi

\bibitem{Xiao2022GQET}
F.~Xiao, Generalized quantum evidence theory, Applied Intelligence (2022) DOI:
  10.1007/s10489--022--04181--0.

\bibitem{2020Quantum}
L.~Pan, X.~Gao, Y.~Deng, Quantum algorithm of dempster combination rule.

\bibitem{wang2022modelling}
Z.~Wang, C.~Mu, S.~Hu, C.~Chu, X.~Li, Modelling the dynamics of regret
  minimization in large agent populations: a master equation approach, in:
  Proceedings of the 31st International Joint Conference on Artificial
  Intelligence (IJCAI-22), 2022, pp. 534--540.

\bibitem{li2021multisource}
D.~Li, Y.~Deng, K.~H. Cheong, Multisource basic probability assignment fusion
  based on information quality, International Journal of Intelligent Systems
  36~(4) (2021) 1851--1875.

\bibitem{liu2022consistency}
P.~Liu, Y.~Li, P.~Wang, {Consistency threshold-and score function-based
  multi-attribute decision-making with Q-rung orthopair fuzzy preference
  relations}, Information Sciences 618 (2022) 356--378.

\bibitem{chang2021transparent}
L.~Chang, L.~Zhang, C.~Fu, Y.-W. Chen, Transparent digital twin for output
  control using belief rule base, IEEE Transactions on Cybernetics (2021) DOI:
  10.1109/TCYB.2021.3063285.

\bibitem{yager2011fusion}
R.~R. Yager, On the fusion of imprecise uncertainty measures using belief
  structures, Information Sciences 181~(15) (2011) 3199--3209.

\bibitem{yager2019uncertain}
R.~R. Yager, N.~Alajlan, Y.~Bazi, Uncertain database retrieval with
  measure-based belief function attribute values, Information Sciences 501
  (2019) 761--770.

\bibitem{Xiao2022NQMF}
F.~Xiao, W.~Pedrycz, Negation of the quantum mass function for multisource
  quantum information fusion with its application to pattern classification,
  IEEE Transactions on Pattern Analysis and Machine Intelligence (2022) DOI:
  10.1109/TPAMI.2022.3167045.

\bibitem{Xiao2022Acomplexweighted}
F.~Xiao, Z.~Cao, C.-T. Lin, A complex weighted discounting multisource
  information fusion with its application in pattern classification, IEEE
  Transactions on Knowledge and Data Engineering (2022) DOI:
  10.1109/TKDE.2022.3206871.

\bibitem{Shang2021compound}
Q.~Shang, H.~Li, Y.~Deng, K.~H. Cheong, {Compound credibility for conflicting
  evidence combination: an autoencoder-K-Means approach}, IEEE Transactions on
  Systems, Man, and Cybernetics: Systems (2021) 10.1109/TSMC.2021.3130187\href
  {https://doi.org/{10.1109/TSMC.2021.3130187}}
  {\path{doi:{10.1109/TSMC.2021.3130187}}}.

\end{thebibliography}

%% else use the following coding to input the bibitems directly in the
%% TeX file.

%\begin{thebibliography}{00}
%
%%% \bibitem[Author(year)]{label}
%%% Text of bibliographic item
%
%\bibitem[()]{}
%
%\end{thebibliography}

\section*{Acknowledgment}
This research is supported by the National Natural Science Foundation of China (No. 62003280), Chongqing Talents: Exceptional Young Talents Project (No. cstc2022ycjh-bgzxm0070), Natural Science Foundation of Chongqing, China (No. CSTB2022NSCQ-MSX0531), and Chongqing Overseas Scholars Innovation Program (No. cx2022024).

\section*{Reference}
\bibliographystyle{elsarticle-num}

\end{document}